# Discussions of gas power cycle performance analysis method in the course of Engineering Thermodynamics

Di He, Zhipeng Duan, Linbo Yan, Chaojun Wang, Boshu He[*]

Institute of Combustion and Thermal Systems, School of Mechanical, Electronic and Control Engineering, Beijing Jiaotong University, Beijing 100044, China

**Abstract:** Engineering Thermodynamics has been the core course of many science and engineering majors at home and abroad, including energy and power, mechanical engineering, civil engineering, aerospace, cryogenic refrigeration, food engineering, chemical engineering, and environmental engineering, among which gas power cycle is one of the important contents. However, many Engineering Thermodynamics textbooks at home and abroad focus only on evaluating the thermal efficiency of gas power cycle, while the important concept of specific cycle net work is ignored. Taking an ideal Otto cycle and an ideal Brayton as examples, the optimum compression ratio (or the pressure ratio) and the maximum specific cycle net work are analyzed and determined. The ideal Otto and the ideal Brayton cycles, and also other gas power cycles, are concluded that the operation under the optimum compression/pressure ratio of the engine, instead of under the higher efficiency, is more economic and more reasonable. We concluded that the two very important concepts, i.e., the maximum specific cycle net work and the optimum compression (or pressure) ratio for the gas power cycles, should be emphasized in the Engineering Thermodynamics teaching process and the latter revised or the newly edited textbooks, in order to better guide the engineering applications. In the end, general *T-s* diagram is proposed for the gas power cycles.

# 1, Introduction

Çengel et al [1-4] wrote in the Preface of the textbooks: Thermodynamics is an exciting and fascinating subject that deals with energy, and thermodynamics has long been an essential part of engineering curricula all over the world. It has a broad application area ranging from microscopic organisms to common household appliances, transportation vehicles, power generation systems, and even philosophy. This implies this curriculum is essential for the engineering discipline.

---

[*] Corresponding author. Email: hebs@bjtu.edu.cn

Moran et al [5-7] wrote in their textbooks that: Engineers use principles drawn from thermodynamics and other engineering sciences, including fluid mechanics and heat and mass transfer, to analyze and design devices intended to meet human needs. Throughout the twentieth century, engineering applications of thermodynamics helped pave the way for significant improvements in our quality of life with advances in major areas such as surface transportation, air travel, space flight, electricity generation and transmission, building heating and cooling, and improved medical practices. In the twenty-first century, engineers will create the technology needed to achieve a sustainable future. Thermodynamics will continue to advance human well-being by addressing looming societal challenges owing to declining supplies of energy resources: oil, natural gas, coal, and fissionable material; effects of global climate change; and burgeoning population. This also means this curriculum is essential for the engineering discipline.

That is to say that Engineering Thermodynamics is quite important both for science and engineering. It is a science that studies the principles of the conversion and the application among thermal energy and mechanical energy. One of the main tasks of Engineering Thermodynamics is to clarify theoretically the ways to improve the efficiency of heat engine and the utilization rate of heat energy.

Along with the progress of science and technology and production development, the scope of the research and application of Engineering Thermodynamics is not limited to just as a heat engine (or refrigeration) theoretical basis. It has now expanded to many fields of engineering technology, such as aerospace, high-energy laser, heat pump, air separation, air conditioning, water desalination, chemical refining, biological engineering, low temperature superconducting, physical chemistry, etc. All the fields need the basic theory of Engineering Thermodynamics and the guidance of the basic knowledge. Therefore, Engineering Thermodynamics has become a compulsory basic course for many related majors.

From the perspective of guiding engineering application, Engineering Thermodynamics plays an important theoretical guiding role in guiding the application related to thermal energy and a very important role in the design and transformation of equipment, and has the significance of guiding light for the efficient use of energy. It is no exaggeration to say that without the correct theoretical system of Engineering Thermodynamics, there would be no modern advanced power equipment.

In this regard, Engineering Thermodynamics textbooks at home and abroad have played a very important role. Currently, typical textbooks used by high-level universities include [1-13], represented by textbooks edited by Cengel et al [1-4], Moran et al [5-7] and Borgnakke et al [8-10], which are basically updated every 4 years. The latest textbook editions are the 9th edition by Cengel et al in 2019 [4], the 9th edition by Moran et al in 2018 [7] and the 10th edition by Borgnakke et al in 2019 [10].

For the performance analyses of gas power cycles, textbooks from home and abroad mainly introduce the composition, performance parameters and performance analysis method for ideal Otto cycle, Diesel cycle, Dual cycle and Brayton cycle. The change rule of performance parameters, such as thermal efficiency, with relevant parameters is presented. However, all textbooks, even monographs [14,15], do not clearly point out: with certain conditions, such as the given cycle operating temperature limit, namely, inlet temperature and maximum cycle temperature, under what state should gas power cycle work most economically? When this part of knowledge was taught in the class, these questions will be naturally raised: is the gas power cycle device with highest thermal efficiency the most economical one? What performance parameters should be used to design or evaluate gas power cycle devices in engineering applications?

As for a gas power cycle device within the given temperature limit, there exists a maximum cycle specific work, corresponding to an optimal compression ratio. The answer can not be found in the textbooks and will be presented in this work. All the textbooks in Chinese and the textbooks in English available do not systematically analyze the piston gas power cycle, and only some textbooks involve in the analysis of Brayton cycle device. For example, the textbook [16] involves in the analysis of gas turbine ideal cycle. In the analysis of Brayton cycle device, the textbook [17] analyzes the conditions with the highest cycle thermal efficiency with an example, and gives the optimal compressor pressure ratio when the cycle specific work is the maximum. The textbooks [1-4] also involve in the analysis of ideal cycle of gas turbine, and give the optimal pressure ratio when the cycle specific power is maximum. The textbooks [5-7] also involve in the analysis of ideal cycle of gas turbine, and an example is used to derive the optimal pressure ratio when the cycle specific power is maximum. Wang made a detailed analysis in his monograph [18] on the most economical condition under which the gas power cycle device should be operated. However, the analyses of the optimal compression ratios are not presented by Wang [18] for the piston devices with irreversible compression and expansion processes, and analyses for advanced efficient piston power cycles, such as Atkinson cycle and Miller cycle, are also not involved. This part should be introduced into the teaching process of Engineering Thermodynamics, and even be introduced into the textbooks of Engineering Thermodynamics. With this knowledge introduced, engineering applications can be better and clearer theoretically guided.

Instructors who are very familiar with the Engineering Thermodynamics course are very clear: for any gas power cycle, as long as the infinitely increased pressure ratio, or compression ratio, namely the infinitely increased pressure or temperature at the end of the compression process, the theoretical maximum thermal efficiency can be infinitely close to the thermal efficiency of Carnot cycle operating within the same temperature limit. Engine designed according to this guiding ideology has very high thermal efficiency, but the specific net work for an engine cycle is infinitesimal, close to zero and this engine is useless. This is clearly not the goal to be pursued in engineering applications of gas power cycle devices, especially the moveable units. This implies that the traditional analysis method of theory guidance for the engineering application problem should be improved a lot.

In this paper, the authors concluded that it is necessary to present directly the performance parameters to evaluate correctly the gas power cycle devices in the textbook of Engineering Thermodynamics. That is to say, ideal gas power cycle device should work at the condition of maximum specific cycle net work instead of maximum thermal efficiency. The actual gas power cycle device has the maximum specific cycle net work condition and the highest cycle thermal efficiency condition. The two conditions are not coincidental. The working condition depends on the nature (movable or stationary) of the device and its purpose. Textbook, organizing and planning the content with these presentations, can provide correct guidance to engineering applications.

Taking the ideal gasoline engine cycle, Otto cycle, and the ideal gas turbine cycle, Brayton cycle, as examples, this paper illustrates the most economical theoretical basis for the gas power cycle device to work at the maximum specific cycle net work and the conditions for the maximum specific cycle net work.

# 2, Otto cycle

The performance of reciprocating engines, such as Otto cycle, Diesel cycle, Dual cycle, Atkinson cycle and Miller cycle engines, can be determined by applying the closed system energy balance and the second law along with property data including mean effective pressure, thermal efficiency, and the effects of varying compression ratio.

The cycle for spark-ignition reciprocating engines is idealized as the Otto cycle in the textbook of Engineering Thermodynamics. The cycle is shown in Fig. 1 (*p-v* diagram) and Fig. 2 (*T-s* diagram), including four reversible processes, i.e., 1-2 Isentropic compression, 2-3 Constant-volume heat addition, 3-4 Isentropic expansion, and 4-1 Constant-volume heat rejection.

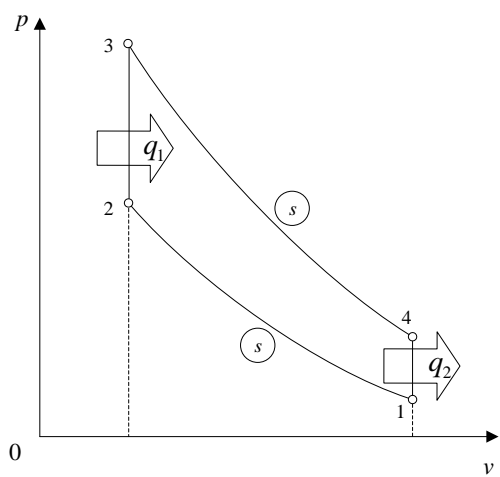
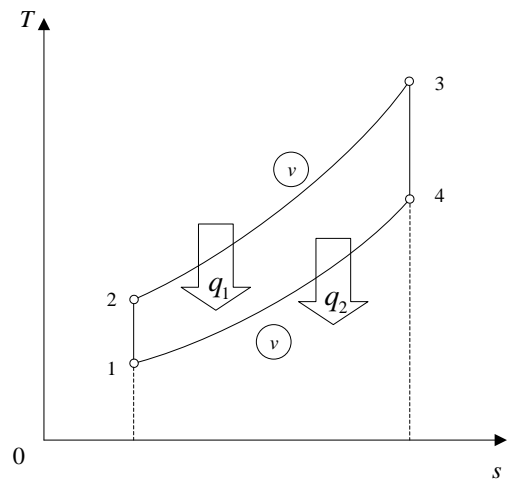

Fig.1 *p-v* diagram for Otto cycle　　　　　Fig.2 *T-s* diagram for Otto cycle

The cold-air-standard assumptions are used in the textbook of Engineering Thermodynamics and also used in this work. At this case, the properties of the working fluids [4] are, the gas constant $R$=0.2870 kJ/kg·K, the specific heat at constant pressure $c_p$=1.005 kJ/kg·K, the specific heat at constant volume $c_V$=0.718 kJ/kg·K and the specific heat ratio $k$=1.400. When the compression ratio, $r = \frac{v_1}{v_2} = \left(\frac{T_2}{T_1}\right)^{\frac{1}{k-1}}$, and the maximum-to-minimum temperature ratio, $\tau = \frac{T_3}{T_1}$, are known, the Otto cycle is then determined. The thermal efficiency of an ideal Otto cycle is

$$\eta_{t,\text{Otto}} = 1 - \frac{1}{r^{k-1}} \qquad (1)$$

As can be seen from Eq. (1), the theoretical thermal efficiency of an ideal Otto cycle can be improved by increasing the compression ratio. At the condition of the determined maximum temperature in the cycle, the temperature of point 3 shown in Fig. 2, if the compression ratio is high enough (of course, it is restricted by various conditions in engineering, but it can be so imaged), the temperature of point 2, the end state of the compression, approaches to that of point 3, and that of point 4, the end state of the expansion, approaches to that of point 1. The Otto cycle will approach, or be very close to the Carnot cycle. The thermal efficiency of the Otto cycle engine is high enough, approaching to or being very close to the thermal efficiency of Carnot cycle working between the same temperature limit. Is this a good thing? Or should the future engineering applications work toward this direction? The current Engineering Thermodynamics textbooks do not theoretically answer this question positively, and students will naturally have such concerns when learning this part of knowledge.

Carefully analyzing the Otto cycle shown in Fig. 2, when the temperature limit of the cycle, i.e., the maximum cycle temperature $T_3$ and the minimum temperature $T_1$, is known, there exists a maximum specific cycle net work corresponding to an optimal compression ratio. The *T-s* diagram of three Otto cycles with different compression ratios between $T_1$ and $T_3$ is shown in Fig. 3. The compression ratio of cycle 1→2'→3'→4'→1 is very small. It is not difficult to see that the thermal efficiency of this cycle is low, and the specific cycle net work, i.e., the area enclosed by the process line, is also very small. The compression ratio of cycle 1→2"→3"→4"→1 is quite large, with very high thermal efficiency (the limit is the efficiency of Carnot cycle in the same temperature limit) on the one hand, but the specific cycle net work is very small (the limit is the net work done, or the cycle specific work is zero) on the other hand. The compression ratio of cycle 1→2→3→4→1 is in the middle, and the thermal efficiency is in the middle, but the specific cycle net work is relatively large. That is, with the change of compression ratio, there must exist a maximum specific cycle net work between $T_1$ and $T_3$. Unfortunately, discussions about the maximum specific net work for an ideal Otto cycle between the determined values of $T_1$ and $T_3$ can not be found in any textbooks of Engineering Thermodynamics and the discussions will be presented in this work.

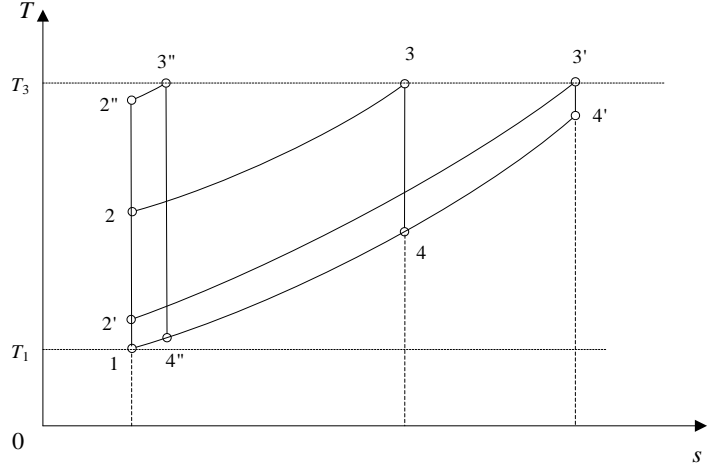

Fig. 3 *T-s* diagram of three Otto cycles with different compression ratios between $T_1$ and $T_3$

**2.1 Optimal compression ratio**

The specific net work for the Otto cycle, shown in Figs. 2 and 3, can be expressed as

$$w_{net} = q_1 - q_2 = c_V(T_3 - T_2) - c_V(T_4 - T_1) = c_V(T_3 - T_2 - T_4 + T_1) \tag{2}$$

and the temperature of each state can be expressed as

$$T_2 = T_1\left(\frac{v_1}{v_2}\right)^{k-1} = T_1 r^{k-1}, \quad T_3 = \tau T_1, \quad T_4 = T_3\left(\frac{v_3}{v_4}\right)^{k-1} = \frac{T_1 T_3}{T_2}$$

Substituting these equations into the specific net work relation, Eq. (2), with the determined values of $T_1$ and $T_3$, and differentiating give

$$\frac{dw_{net}}{dT_2} = c_V\left(-1 + \frac{T_1 T_3}{T_2^2}\right) \tag{3}$$

$$\frac{d^2 w_{net}}{dT_2^2} = -2c_V \frac{T_1 T_3}{T_2^3} < 0 \tag{4}$$

Therefore, the specific cycle net work does have a maximum value. Let Eq. (3) be zero, then the optimal end point temperature of compression is

$$T_{2,opt} = \sqrt{T_1 T_3} \tag{5}$$

Accordingly, when the specific cycle net work is the maximum, the optimal end point temperature of expansion is

$$T_{4,opt} = \sqrt{T_1 T_3} = T_{2,opt} \tag{6}$$

An important conclusion is drawn based on Eq. (6): for an ideal Otto cycle, when the compression end point temperature, $T_2$, is exactly equal to the expansion end point temperature, $T_4$, the specific cycle net work is the maximum value. In other words, when $T_2 = T_4$, the maximum specific cycle net work is reached for an Otto cycle. At this time, the optimal compression ratio is:

$$r_{opt}^{k-1} = \frac{\sqrt{T_1 T_3}}{T_1} = \sqrt{\tau}$$

or

$$r_{opt} = \tau^{\frac{1}{2(k-1)}} \qquad (7)$$

The reason that the maximum value of specific cycle net work varies with the change of compression ratio is:

1) when the compression ratio is less than $r_{opt}$, the working fluid temperature of end point of expansion is always higher than that of the end point of compression, i.e., $T_4 > T_2$. In this case, the increase of $q_1$ (the heat transfer to the working fluid) with the increment of $r$ is larger than that of $q_2$ (the heat transfer from the working fluid) with the increment of $r$, and the specific cycle net work, $w_{net}$, increases with the increment of $r$.

2) when the compression ratio is larger than $r_{opt}$, the temperature of end point of expansion is always lower than that of the end point of compression, i.e., $T_4 < T_2$. In this case, the increase of $q_1$ with the increment of $r$ is smaller than that of $q_2$, and the specific cycle net work, $w_{net}$, decreases with the increment of $r$.

3) when the compression ratio is equal to $r_{opt}$, the temperature of end point of expansion is equal to that of the end point of compression, i.e., $T_4 = T_2$. In this case, the increase of $q_1$ with the increment of $r$ is equal to that of $q_2$, and the specific cycle net work, $w_{net}$, reaches its maximum.

## 2.2 Maximum specific cycle net work

Substituting Eq. (5) for the temperature of the optimal end point of compression, and Eq. (6) for the temperature of the optimal end point of expansion equations into Eq. (2), the specific cycle net work relation for the Otto cycle and simplifying give

$$w_{net,max} = c_V \left( \sqrt{T_3} - \sqrt{T_1} \right)^2$$

Or the dimensionless maximum specific cycle net work

$$\frac{w_{net,max}}{c_V T_1} = \left( \sqrt{\frac{T_3}{T_1}} - 1 \right)^2 = \left( \sqrt{\tau} - 1 \right)^2 \qquad (8)$$

This means that the dimensionless maximum specific cycle net work is determined only by the temperature ratio, or, the maximum specific cycle net work of an ideal Otto internal combustion engine is only a function of the specific heat at constant volume of the working fluid, the highest and lowest operating temperatures, and is independent of the specific heat ratio. The maximum specific cycle net work can be increased by raising the maximum operating temperature, lowering the minimum working fluid temperature and choosing the working fluid with larger specific heat at constant volume.

**2.3 Optimal thermal efficiency**

The optimal thermal efficiency corresponds to the optimal compression ratio at which the maximum specific cycle net work is reached. Substituting the optimal compression ratio equation, Eq. (7), into the thermal efficiency relation, Eq. (1), and simplifying give the optimal thermal efficiency

$$\eta_{t,\text{Otto,opt}} = 1 - \frac{1}{\sqrt{\tau}} \tag{9}$$

This means the optimal thermal efficiency is only the function of temperature ratio, $\tau=T_3/T_1$, independent of working fluid properties, and increases with the increase of $T_3/T_1$. That is, increasing $T_3$ or decreasing $T_1$ can improve the Otto engine thermal efficiency, which is consistent with the Carnot corollaries [5-7].

The analysis also shows that the specific cycle net work of the Otto engine is inconsistent with its thermal efficiency, one of the traditional economic parameter. When the specific cycle net work is the maximum, the thermal efficiency does not reach its maximum value. When the thermal efficiency approaches or reaches its maximum value, the specific cycle net work approaches zero.

**2.4 Examples**

A simple ideal Otto cycle is taken as an example. The analysis shows the performance parameters of the cycle operating at a given compression ratio or an optimal compression ratio under given conditions, such as heat source temperature and environment temperature, especially the changes of specific cycle net work, the thermal efficiency, the exergy destructions, and the second-law efficiency of this cycle.

**Example 1** An ideal air-standard Otto cycle has a compression ratio of 8. At the beginning of the compression process, air is at 100 kPa and 300 K, and 800 kJ/kg of heat is transferred to air during the constant-volume heat-addition process from a source at 1900 K and waste heat is rejected to the surroundings at 300 K. Determine (a) the maximum temperature and pressure that occur during the cycle, (b) the specific net work and the thermal efficiency, (c) the mean effective pressure for the cycle, (d) the exergy destruction associated with each of the four processes and the cycle and (e) the second-law efficiency of this cycle.

**Assumptions 1** The air-standard assumptions are applicable. **2** Kinetic and potential energy changes are negligible. **3** Steady operating conditions exist.

**Analysis** The $p$-$v$ and $T$-$s$ diagrams of the ideal Otto cycle described is shown in Figs. 1 and 2. We note that the air contained in the cylinder forms a closed system.

(a) The maximum temperature and pressure in an Otto cycle occur at the end of the constant-volume heat-addition process (state 3). But first we need to determine the temperature and pressure of air at the end of the isentropic compression process (state 2) according to the air-standard assumptions and the process relations.

State 1
$$p_1 = 0.10 \text{ MPa}, \quad T_1 = 300 \text{ K}$$
$$v_1 = \frac{RT_1}{p_1} = \frac{287.0 \times 300}{0.10 \times 10^6} = 0.8610 \text{ m}^3/\text{kg}$$

State 2(isentropic compression of an ideal gas from state 1)
$$v_2 = \frac{v_1}{r} = \frac{0.8610}{8} = 0.1076 \text{ m}^3/\text{kg}$$
$$T_2 = T_1 r^{k-1} = 300 \text{ K} \times 8^{1.4-1} = 689.22 \text{ K}$$
$$p_2 = p_1 \left(\frac{v_1}{v_2}\right)^k = p_1 r^k = 0.10 \times 8^{1.4} = 1.8379 \text{ MPa}$$

State 3(constant-volume heat addition from state 2)
$$v_3 = v_2 = 0.1076 \text{ m}^3/\text{kg}$$
$$q_1 = c_V (T_3 - T_2) = 0.718 \times (T_3 - 689.22) = 800 \text{ kJ/kg}$$

Thus
$$T_3 = 1803.42 \text{ K}$$
$$p_3 = p_2 \frac{T_3}{T_2} = 1.8379 \times \frac{1803.42}{689.22} = 4.8091 \text{ MPa}$$

State 4(isentropic expansion of an ideal gas from state 3)
$$v_4 = v_1 = 0.8610 \text{ m}^3/\text{kg}$$
$$p_4 = p_3 \left(\frac{v_3}{v_4}\right)^k = 4.8091 \times \left(\frac{0.1076}{0.8610}\right)^{1.4} = 4.8091 \times \left(\frac{1}{8}\right)^{1.4} = 0.2617 \text{ MPa}$$
$$T_4 = \frac{p_4 v_4}{R} = \frac{0.2617 \times 10^6 \times 0.8610}{287.0} = 784.99 \text{ K}$$

(b) the specific net work and the thermal efficiency

The heat transfer from the working fluid is
$$q_2 = c_V (T_4 - T_1) = 0.718 \times (784.99 - 300) = 348.22 \text{ kJ/kg}$$

The specific net work for one cycle is
$$w_{\text{net}} = q_1 - q_2 = 800 - 348.22 = 451.78 \text{ kJ/kg}$$

The thermal efficiency of the cycle is determined from its definition
$$\eta_{\text{th}} = \frac{w_{\text{net}}}{q_1} = \frac{451.78}{800} = 0.5647 \quad \text{or} \quad 56.47\%$$

Under the cold-air-standard assumptions, the thermal efficiency would be
$$\eta_{t,\text{Otto}} = 1 - \frac{1}{r^{k-1}} = 1 - \frac{1}{8^{1.4-1}} = 0.5647 \quad \text{or} \quad 56.47\%$$

(c) The mean effective pressure is determined from its definition

$$MEP = \frac{w_{net}}{v_1 - v_2} = \frac{451.78}{0.8610 - 0.1076} = 599.67 \text{ kPa} = 0.60 \text{ MPa}$$

(d) The exergy destructions

Processes 1-2 and 3-4 are isentropic ($s_1=s_2$, $s_3=s_4$) and therefore do not involve any internal or external irreversibilities; that is, $X_{dest,12} = 0$ and $X_{dest,34} = 0$.

Processes 2-3 and 4-1 are constant-volume heat-addition and heat-rejection processes, respectively, and are internally reversible. However, the heat transfer between the working fluid and the source or the sink takes place through a finite temperature difference, rendering both processes irreversible. The exergy destruction associated with each process is determined. However, first we need to determine the entropy change of air during these processes:

The entropy change for process 2-3 of ideal gas

$$s_3 - s_2 = c_{V0} \ln \frac{T_3}{T_2}$$

$$= 0.718 \text{ kJ/kg} \cdot \text{K} \times \ln \frac{1803.42 \text{ K}}{689.22 \text{ K}} = 0.6906 \text{ kJ/kg} \cdot \text{K}$$

Also

$q_1$=1000 kJ/kg and $T_H$=1900 K

Thus

$$x_{dest,23} = T_0 \left[ (s_3 - s_2)_{sys} - \frac{q_1}{T_H} \right]$$

$$= 300 \text{ K} \times \left( 0.6906 \text{ kJ/kg} \cdot \text{K} - \frac{800 \text{ kJ/kg}}{1900 \text{ K}} \right) = 80.87 \text{ kJ/kg}$$

The average temperature of air in the constant-volume heat-addition is

$$\bar{T}_1 = \frac{q_1}{s_3 - s_2} = \frac{800 \text{ kJ/kg}}{0.6906 \text{ kJ/kg} \cdot \text{K}} = 1158.36 \text{ K}$$

For process 4-1, $s_4 - s_1 = s_2 - s_3 = -0.6906 \text{ kJ/kg} \cdot \text{K}$, $q_{41} = q_2 = 348.22 \text{ kJ/kg}$, and $T_{sink} = 300 \text{ K}$. Thus,

$$x_{dest,\,41} = T_0 \left[ (s_1 - s_4)_{sys} + \frac{q_{41}}{T_{sink}} \right]$$

$$= 300 \text{ K} \times \left[ -0.6906 \text{ kJ/kg} \cdot \text{K} + \frac{348.22 \text{ kJ/kg}}{300 \text{ K}} \right] = 141.03 \text{ kJ/kg}$$

Therefore, the irreversibility of the cycle is

$$x_{dest,cycle} = x_{dest,12} + x_{dest,23} + x_{dest,34} + x_{dest,41}$$
$$= 0 + 80.87 \text{ kJ/kg} + 0 + 141.03 \text{ kJ/kg} = 221.90 \text{ kJ/kg}$$

Notice that the largest exergy destruction in the cycle occurs during the heat-rejection process. Therefore, any attempt to reduce the exergy destruction, or improve the performance, should start with this process.

The average temperature of air in the constant-volume heat-rejection is

$$\overline{T}_2 = \frac{q_2}{s_3 - s_2} = \frac{348.22 \text{ kJ/kg}}{0.6906 \text{ kJ/kg} \cdot \text{K}} = 504.21 \text{ K}$$

The thermal efficiency of the cycle is determined from the average temperature

$$\eta_{t,\text{Otto}} = 1 - \frac{\overline{T}_2}{\overline{T}_1} = 1 - \frac{504.21 \text{ K}}{1158.36 \text{ K}} = 0.5647 \quad \text{or } 56.47\%$$

(e) The second-law efficiency is defined as

$$\eta_{\text{II}} = \frac{\text{Exergy recovered}}{\text{Exergy expended}} = \frac{x_{\text{recovered}}}{x_{\text{expended}}} = 1 - \frac{x_{\text{destroyed}}}{x_{\text{expended}}}$$

Here the expended energy is the energy content of the heat supplied to the air in the engine (which is its work potential) and the energy recovered is the net work output

$$x_{\text{expended}} = x_{\text{heat,in}} = \left(1 - \frac{T_0}{T_H}\right) q_{\text{in}}$$

$$= \left(1 - \frac{300 \text{ K}}{1900 \text{ K}}\right) \times 800 \text{ kJ/kg} = 673.68 \text{ kJ/kg}$$

$$x_{\text{recovered}} = w_{\text{net}} = 451.78 \text{ kJ/kg}$$

Substituting, the second-law efficiency of this cycle is determined to be

$$\eta_{\text{II}} = \frac{x_{\text{recovered}}}{x_{\text{expended}}} = \frac{451.78 \text{ kJ/kg}}{673.68 \text{ kJ/kg}} = 0.6706 \quad \text{or } 67.06\%$$

or

$$\eta_{\text{II}} = 1 - \frac{x_{\text{destroyed}}}{x_{\text{expended}}} = 1 - \frac{221.90 \text{ kJ/kg}}{673.68 \text{ kJ/kg}} = 0.6706 \quad \text{or } 67.06\%$$

**Example 2** If the ideal air-standard Otto cycle operates between the lowest temperature $T_1$=300 K, and $T_3$=1803.42 K, and at the compression ratio of optimal instead of 8, repeat Example 1.

**Assumptions 1** The air-standard assumptions are applicable. **2** Kinetic and potential energy changes are negligible. **3** Steady operating conditions exist.

**Analysis** The *p-v* and *T-s* diagrams of the ideal Otto cycle described is shown in Figs. 1 and 2. We note that the air contained in the cylinder forms a closed system. The optimal compression ratio should be firstly determined with the given temperature limit.

From Eq. (7), the optimal compression ratio is

$$r_{\text{opt}} = \left(\frac{T_3}{T_1}\right)^{\frac{1}{2(k-1)}} = \left(\frac{1803.42}{300}\right)^{\frac{1}{2 \times (1.4-1)}} = 9.4128$$

(a) The maximum temperature and pressure in an Otto cycle occur at the end of the constant-volume heat-addition process (state 3). The temperature is known at state 3. The pressure of air at the end of the isentropic compression process (state 2) can be determined according to the air-standard assumptions and the process relations.

State 1

$$p_1 = 0.10 \text{ MPa}, \quad T_1 = 300 \text{ K}, \quad v_1 = 0.8610 \text{ m}^3/\text{kg}$$

State 2(isentropic compression of an ideal gas from state 1)

$$v_2 = \frac{v_1}{r_{opt}} = \frac{0.8610}{9.4128} = 0.09147 \text{ m}^3/\text{kg}$$

$$T_2 = T_1 r_{opt}^{k-1} = 300 \text{ K} \times 9.1428^{1.4-1} = 735.55 \text{ K}$$

$$p_2 = p_1 \left(\frac{v_1}{v_2}\right)^k = p_1 r^k = 0.10 \times 9.4128^{1.4} = 2.3078 \text{ MPa}$$

State 3(constant-volume heat addition from state 2)

$$v_3 = v_2 = 0.09147 \text{ m}^3/\text{kg}$$

$$q_1 = c_V (T_3 - T_2) = 0.718 \times (1803.42 - 735.54) = 766.74 \text{ kJ/kg}$$

$$p_3 = p_2 \frac{T_3}{T_2} = 2.3078 \times \frac{1803.42}{735.54} = 5.6584 \text{ MPa}$$

State 4(isentropic expansion of an ideal gas from state 3)

$$v_4 = v_1 = 0.8610 \text{ m}^3/\text{kg}$$

$$p_4 = p_3 \left(\frac{v_3}{v_4}\right)^k = 5.6584 \times \left(\frac{1}{9.1248}\right)^{1.4} = 0.2452 \text{ MPa}$$

$$T_4 = \frac{p_4 v_4}{R} = \frac{0.2452 \times 10^6 \times 0.8610}{287.0} = 735.55 \text{ K} = T_2$$

(b) the specific net work and the thermal efficiency

The heat transfer from the working fluid is

$$q_2 = c_V (T_4 - T_1) = 0.718 \times (735.55 - 300) = 312.72 \text{ kJ/kg}$$

The specific net work for one cycle is

$$w_{net} = q_1 - q_2 = 766.74 - 312.72 = 454.02 \text{ kJ/kg}$$

The thermal efficiency of the cycle is determined from its definition

$$\eta_{th} = \frac{w_{net}}{q_1} = \frac{454.02}{766.74} = 0.5921 \text{ or } 59.21\%$$

Under the cold-air-standard assumptions, the thermal efficiency would be

$$\eta_{t,Otto} = 1 - \frac{1}{r^{k-1}} = 1 - \frac{1}{9.4128^{1.4-1}} = 0.5921 \text{ or } 59.21\%$$

(c) The mean effective pressure is determined from its definition

$$MEP = \frac{w_{net}}{v_1 - v_2} = \frac{454.02}{0.8610 - 0.09147} = 590.0 \text{ kPa} = 0.59 \text{ MPa}$$

(d) The exergy destructions

The exergy destructions can be determined following the ways shown in example 1.

The entropy change for process 2-3 of ideal gas is

$$s_3 - s_2 = c_{V0} \ln \frac{T_3}{T_2}$$

$$= 0.718 \text{ kJ/kg} \cdot \text{K} \times \ln \frac{1803.42 \text{ K}}{735.55 \text{ K}} = 0.6439 \text{ kJ/kg} \cdot \text{K}$$

Also

$$q_1 = 766.74 \text{ kJ/kg and } T_H = 1900 \text{ K}$$

Thus

$$x_{\text{dest},23} = T_0 \left[ (s_3 - s_2)_{\text{sys}} - \frac{q_1}{T_H} \right]$$

$$= 300 \text{ K} \times \left( 0.6439 \text{ kJ/kg} \cdot \text{K} - \frac{766.74 \text{ kJ/kg}}{1900 \text{ K}} \right) = 72.11 \text{ kJ/kg}$$

The average temperature of air in the constant-volume heat-addition is

$$\bar{T}_1 = \frac{q_1}{s_3 - s_2} = \frac{766.74 \text{ kJ/kg}}{0.6439 \text{ kJ/kg} \cdot \text{K}} = 1190.78 \text{ K}$$

For process 4-1, $s_4 - s_1 = s_2 - s_3 = -0.6439$ kJ/kg·K, $q_{41} = q_2 = 312.72$ kJ/kg, and $T_{\text{sink}} = 300$ K. Thus,

$$x_{\text{dest},41} = T_0 \left[ (s_1 - s_4)_{\text{sys}} + \frac{q_{41}}{T_{\text{sink}}} \right]$$

$$= 300 \text{ K} \times \left[ -0.6439 \text{ kJ/kg} \cdot \text{K} + \frac{312.72 \text{ kJ/kg}}{300 \text{ K}} \right] = 119.55 \text{ kJ/kg}$$

Therefore, the irreversibility of the cycle is

$$x_{\text{dest,cycle}} = x_{\text{dest},12} + x_{\text{dest},23} + x_{\text{dest},34} + x_{\text{dest},41}$$

$$= 0 + 72.11 \text{ kJ/kg} + 0 + 119.55 \text{ kJ/kg} = 191.66 \text{ kJ/kg}$$

Also, notice that the largest exergy destruction in the cycle occurs during the heat-rejection process. Therefore, any attempt to reduce the exergy destruction, or improve the performance, should start with this process.

The average temperature of air in the constant-volume heat-rejection is

$$\bar{T}_2 = \frac{q_2}{s_3 - s_2} = \frac{312.72 \text{ kJ/kg}}{0.6439 \text{ kJ/kg} \cdot \text{K}} = 485.67 \text{ K}$$

The thermal efficiency of the cycle is determined from the average temperature

$$\eta_{t,\text{Otto}} = 1 - \frac{\bar{T}_2}{\bar{T}_1} = 1 - \frac{485.67 \text{ K}}{1190.78 \text{ K}} = 0.5921 \text{ or } 59.21\%$$

(e) The second-law efficiency is

$$x_{\text{expended}} = \left( 1 - \frac{T_0}{T_H} \right) q_{\text{in}} = \left( 1 - \frac{300 \text{ K}}{1900 \text{ K}} \right) \times 766.74 \text{ kJ/kg} = 645.68 \text{ kJ/kg}$$

$$x_{\text{recovered}} = w_{\text{net}} = 454.02 \text{ kJ/kg}$$

Substituting, the second-law efficiency of this cycle is determined to be

$$\eta_{\mathrm{II}} = \frac{x_{\mathrm{recovered}}}{x_{\mathrm{expended}}} = \frac{454.02 \text{ kJ/kg}}{645.68 \text{ kJ/kg}} = 0.7032 \text{ or } 70.32\%$$

or

$$\eta_{\mathrm{II}} = 1 - \frac{x_{\mathrm{destroyed}}}{x_{\mathrm{expended}}} = 1 - \frac{191.66 \text{ kJ/kg}}{645.68 \text{ kJ/kg}} = 0.7032 \text{ or } 70.32\%$$

Comparing the results from Examples 1 and 2, the Otto engine operates between the lowest temperature $T_1$=300 K and the highest working temperature $T_3$=1803.42 K having a determined optimal compression ratio and calculated by Eq. (7). When the Otto cycle operates at the compression ratio of 8, as shown in example 1, the specific cycle net work, the thermal efficiency and the second-law efficiency are 451.78 kJ/kg, 56.47% and 67.06%, respectively. While, when it operates at the optimal compression ratio of 9.4128, as shown in example 2, the specific cycle net work, the thermal efficiency and the second-law efficiency are 454.02 kJ/kg, 59.21% and 70.32%, respectively. Obviously, the performance of the Otto engine operates at the optimal compression ratio determined by the maximum temperature and the minimum temperature is better than that operates at compression ratio of 8 in the same temperature limit. The Otto cycle operates at the optimal compression ratio does the maximum specific net work. The thermal efficiency and the second-law efficiency are not always increased, although in the examples they are. When the optimal compression ratio is larger than the original compression ratio, the efficiencies at the optimal compression ratio will increase, otherwise, they will decrease.

Working at the optimal compression ratio, even though the efficiency of an ideal Otto cycle is decreased (when $r_{\mathrm{opt}} > r$ it will increase), the specific cycle net work is maximized, and then the equipment can be miniaturized; or, with the device of the same size, the net output work will be larger. That is, the Otto cycle operating at the optimal compression ratio is more economical than that operating at other compression ratios. In other words, when the Otto cycle operates at $r < r_{\mathrm{opt}}$, the overall economy of the device is better as a result of the improvement of optimal compression ratio of the device, resulting the increased thermal efficiency, the specific cycle net work and the exergy efficiency. On the whole, it is more economical and reasonable for the Otto cycle to operate at the optimal compression ratio

## 3, Ideal air-standard Brayton cycle

The two major application areas of gas-turbine engines are aircraft propulsion and electric power generation. At present, in the Engineering Thermodynamics textbooks at home and abroad, a gas turbine engine is simplified into an ideal Brayton cycle consisting of isentropic compression (in a compressor), constant-pressure heat addition, isentropic expansion (in a turbine) and constant-pressure heat rejection. The ideal Brayton cycle can be shown in Figs. 4 and 5. When the pressure ratio, $\pi = p_2/p_1$, and temperature ratio, $\tau = T_3/T_1$, of the cycle are determined, the cycle is then determined.

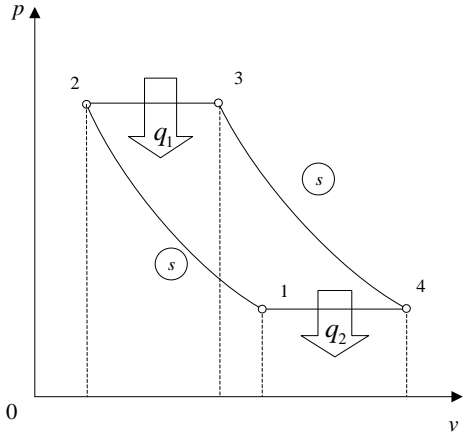
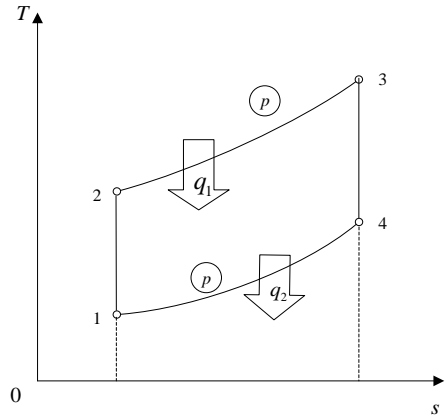

Fig.4 *p-v* diagram of for the ideal Brayton cycle

Fig.5 *T-s* diagram of for the ideal Brayton cycle

When an ideal Brayton cycle is analyzed on a cold air-standard basis, the specific heats are taken as constant. Kinetic and potential energy effects are negligible. The thermal efficiency would be

$$\eta_{t,\text{Brayton}} = 1 - \frac{1}{\pi^{\frac{k-1}{k}}} \tag{10}$$

This means that under the cold-air-standard assumptions, the thermal efficiency of an ideal Brayton cycle depends on the pressure ratio of the gas turbine and the specific heat ratio of the working fluid. The thermal efficiency increases with both of these parameters, which is also the case for actual gas turbines.

Increasing the pressure ratio can largely improve the thermal efficiency until it approaches the thermal efficiency of the Carnot cycle operating in the same temperature limit. However, when the thermal efficiency is high enough, the specific cycle net work is very small, even close to zero, which is not the goal pursued by power devices, especially movable power devices.

Like the Otto cycle shown in Fig. 3, when the temperature limit of an ideal Brayton cycle, i.e., the maximum cycle temperature $T_3$ and the minimum temperature $T_1$, is known, there exists a maximum specific cycle net work corresponding to an optimal pressure ratio and this idea should be known by students in the Engineering Thermodynamics class. However, there are only a few clues for the maximum specific cycle net work of the ideal Brayton cycle which can be found in the textbooks [1-7, 16, 17].

**3.1 Optimal pressure ratio**

Notice that all four processes of the ideal Brayton cycle, as shown in Figs. 4 and 5, are executed in steady-flow devices; thus, they should be analyzed as steady-flow processes. When the changes in kinetic and potential energies are neglected, the specific cycle net work can be expressed as

$$w_{net} = c_p(T_3 - T_2) - c_p(T_4 - T_1) = c_p(T_3 - T_2 - T_4 + T_1)$$

or

$$w_{net} = c_p(T_3 - T_2 - T_1 T_3 / T_2 + T_1) \tag{11}$$

With the determined maximum cycle temperature $T_3$ and the minimum temperature $T_1$, i.e., the determined temperature ratio, $\tau = \frac{T_3}{T_1}$, the specific cycle net work is varied only with $T_2$, the end temperature of the compression depending on the pressure ratio of the cycle. Differentiating Eq. (11) give

$$\frac{dw_{net}}{dT_2} = -1 - \frac{T_1 T_3}{T_2^2} = 0$$

Thus

$$T_{2,\,opt} = \sqrt{T_1 T_3} = T_{4,\,opt} \tag{12}$$

This indicates that the specific net work for a cycle will be maximum when the end temperature of the compression is equal to the end temperature of the expansion. Notice that Eq. (12) for the ideal Brayton cycle is the same for the ideal Otto cycle as shown by Eq. (6). The pressure ratio corresponding to maximum specific cycle net work is the optimal pressure ratio.

$$T_{2,\,opt} = T_1 \pi_{opt}^{\frac{k-1}{k}}$$

Thus

$$\pi_{opt} = \tau^{\frac{k}{2(k-1)}} \tag{13}$$

### 3.2 Maximum specific cycle net work

Substituting Eq. (12) for the temperatures of the optimal end point of compression and the optimal end point of expansion relation into Eq. (11), the specific cycle net work relation for the ideal Brayton cycle and simplifying give

$$w_{net} = c_p \left(\sqrt{T_3} - \sqrt{T_1}\right)^2 \tag{14}$$

Or the dimensionless maximum specific cycle net work

$$\frac{w_{net}}{c_p T_1} = \left(\sqrt{\tau} - 1\right)^2 \tag{15}$$

This means that the dimensionless maximum specific cycle net work is determined only by the temperature ratio, or, the maximum specific cycle net work of an ideal gas-turbine engine is only a function of the specific heat at constant pressure of the working fluid, the highest and the lowest operating temperatures, and is independent of the specific heat ratio. The maximum specific cycle net work can be increased by raising the maximum operating temperature, lowering the minimum working fluid temperature and choosing the working fluid with larger specific heat at constant pressure.

**3.3 Optimal thermal efficiency**

Substituting Eq. (13) for optimal pressure ratio into Eq. (10), for the thermal efficiency of the ideal Brayton cycle and simplifying give

$$\eta_{t,\text{Brayton}} = 1 - \frac{1}{\pi_{\text{opt}}^{\frac{k-1}{k}}} = 1 - \frac{1}{\left(\frac{T_3}{T_1}\right)^{\frac{1}{2}}}$$

Thus

$$\eta_{t,\text{opt}} = 1 - \sqrt{\frac{1}{\tau}} \qquad (15)$$

The optimal thermal efficiency is only determined by the temperature ratio. By the way, the thermal efficiency corresponding to maximum specific cycle net work is not the highest one operating in the same temperature limit.

Therefore, there should be a compromise between the pressure ratio, thus the thermal efficiency, and the net work output according to the application areas of gas-turbine engines. With less work output per cycle, a larger mass flow rate, thus a larger system, is needed to maintain the same power output, which may not be economical.

**3.4 Examples**

A simple ideal Brayton cycle is taken as an example. The analysis shows the performance parameters of the cycle operating at a given pressure ratio or at the optimal pressure ratio under the minimum and the maximum temperatures, especially the changes of specific cycle net work and the thermal efficiency.

**Example 3** An ideal air-standard Brayton cycle has a pressure ratio of 11. At the beginning of the compression process, air is at 100 kPa and 300 K. The maximum cycle temperature is 1300 K. Determine (a) the thermal efficiency and the specific net work, (b) if the net power output is 150 MW, determine the volume flow rate of the air into the compressor, (c) if the compressor operates at the optimal pressure ratio, determine the thermal efficiency, the specific net work and the volume flow rate of the air.

**Assumptions** 1 Steady operating conditions exist. 2 The air-standard assumptions are applicable. 3

Kinetic and potential energy changes are negligible. 4 Air is an ideal gas with constant specific heats.

**Analysis** The *p-v* and *T-s* diagrams of the ideal Brayton cycle described is shown in Figs. 4 and 5. The state parameters of state 1 should be determined first for the estimation of the volume flow rate of the air into the compressor.

$$p_1 = 0.10 \text{ MPa}, \quad T_1 = 300 \text{ K}$$

$$v_1 = \frac{RT_1}{p_1} = \frac{287.0 \times 300}{0.10 \times 10^6} = 0.8610 \text{ m}^3/\text{kg}$$

The temperatures of state 2 and state 4

$$T_2 = T_1 \left(\frac{p_2}{p_1}\right)^{\frac{k-1}{k}} = 300 \text{ K} \times (11)^{\frac{1.4-1}{1.4}} = 595.20 \text{ K}$$

$$T_4 = T_3 \left(\frac{p_4}{p_3}\right)^{\frac{k-1}{k}} = 1300 \text{ K} \times \left(\frac{1}{11}\right)^{\frac{1.4-1}{1.4}} = 655.24 \text{ K}$$

(a) The thermal efficiency for an ideal Brayton cycle can be determined from Eq. (10)

$$\eta_{t,\text{Brayton}} = 1 - \frac{1}{\pi^{\frac{k-1}{k}}} = 1 - \frac{1}{11^{\frac{1.4-1}{1.4}}} = 49.60\%$$

The specific net work output for a cycle from is

$$w_{\text{net}} = c_p (T_3 - T_2 - T_4 + T_1)$$
$$= 1.005 \text{ kJ/kg} \cdot \text{K} \times (1300 \text{ K} - 595.20 \text{ K} - 655.24 \text{ K} + 300 \text{ K})$$
$$= 351.31 \text{ kJ/kg}$$

(b) The volume flow rate of the air into the compressor can be determined as

$$\dot{V} = v_1 \dot{m} = v_1 \frac{\dot{W}_{\text{net}}}{w_{\text{net}}} = 0.8610 \text{ m}^3/\text{kg} \times \frac{150 \text{ MW}}{351.31 \text{ kJ/kg}} = 367.62 \text{ m}^3/\text{s}$$

(c) First, the optimal pressure ratio can be determined from Eq. (13)

$$\pi_{\text{opt}} = \tau^{\frac{k}{2(k-1)}} = \left(\frac{1300}{300}\right)^{\frac{1.4}{2 \times (1.4-1)}} = 13.0148$$

The thermal efficiency for an ideal Brayton cycle at the optimal pressure ratio, then, can be determined from Eq. (10) or Eq. (15)

$$\eta_{t,\text{Brayton}} = 1 - \frac{1}{\pi_{\text{opt}}^{\frac{k-1}{k}}} = 1 - \frac{1}{13.0148^{\frac{1.4-1}{1.4}}} = 0.5196 \text{ or } 51.96\%$$

$$\eta_{t,\text{ opt}} = 1 - \sqrt{\frac{T_1}{T_3}} = 1 - \sqrt{\frac{300 \text{ K}}{1300 \text{ K}}} = 0.5196 \text{ or } 51.96\%$$

The maximum specific net work from Eq. (14) then is

$$w_{net,max} = c_p \left(\sqrt{T_3} - \sqrt{T_1}\right)^2$$
$$= 1.005 \text{ kJ/(kg·K)} \times \left(\sqrt{1300 \text{ K}} - \sqrt{300 \text{ K}}\right)^2$$
$$= 352.76 \text{ kJ/kg}$$

The volume flow rate of the air into the compressor can be determined as

$$\dot{V} = v_1 \dot{m} = v_1 \frac{\dot{W}_{net}}{w_{net,max}} = 0.8610 \text{ m}^3/\text{kg} \times \frac{150 \text{ MW}}{352.76 \text{ kJ/kg}} = 366.11 \text{ m}^3/\text{s}$$

**Discussion:** The specific cycle net work reaches the maximum value when an ideal air-standard Brayton cycle operates at the optimal pressure ratio under given conditions. In this example, since the optimal pressure ratio is larger than the original operating pressure ratio (but the change is not significant, so the specific cycle net work increases only a little), the thermal efficiency of the cycle is also improved, i.e., $\eta_{t,\,opt} = 0.5196 > \eta_{t,Brayton} = 0.4960$. After working at the optimal pressure ratio, the increase of specific cycle net work is inevitable, but the improvement of efficiency is not certain depending on the pressure ratio.

## 4, General *T-s* diagram

Following Fig. 3, general *T-s* diagram is proposed for all the gas power cycles operating between $T_1$ and $T_3$, as shown in Fig. 6. Just as discussed above, all the ideal gas power cycles have maximum specific cycle net work between $T_1$ and $T_3$.

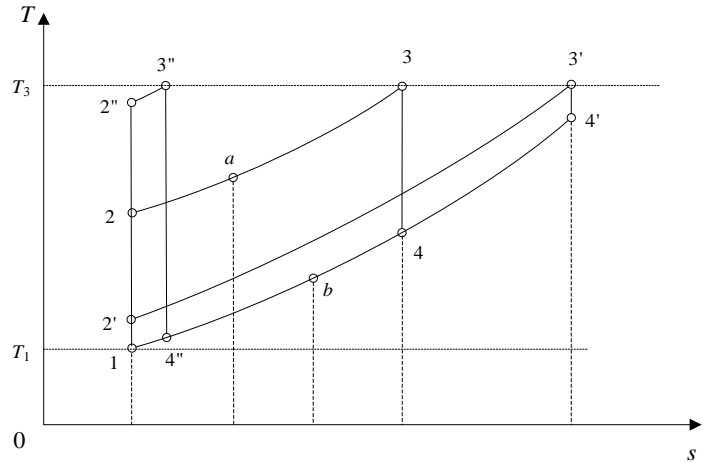

Fig. 6 General *T-s* diagram for the ideal gas power cycles with different compression/pressure ratios between $T_1$ and $T_3$

Taking the cycle of 12a34b1, or 12341 shown in Fig.6, as the example. All the ideal gas power cycles experience the same processes of isentropic compression (1-2) and isentropic expansion (3-4). Processes of different heat addition and heat rejection make the cycle different.

Otto cycle: the heat addition process, 2-*a*-3, is under constant-volume and the heat rejection process, 4-*b*-1, is under constant-volume, as shown in Fig. 2.

Atkinson cycle: the heat addition process, 2-*a*-3, is under constant-volume and the heat rejection process, 4-*b*-1, is under constant-pressure.

Miller-Otto cycle: the heat addition process, 2-*a*-3, is under constant-volume and the heat rejection process, 4-*b*-1, includes two heat rejection processes. The heat rejection process, 4-*b*, is under constant-volume and the other heat rejection process, *b*-1, is under constant-pressure.

Dual cycle: the heat addition process, 2-*a*-3, includes two heat addition processes. The heat addition process, 2-*a*, is under constant-volume and the other heat addition process, *a*-3, is under constant-pressure. The heat rejection process, 4-*b*-1, is under constant-volume.

Diesel cycle: the heat addition process, 2-*a*-3, is under constant-pressure and the heat rejection process, 4-*b*-1, is under constant-volume.

Miller-Diesel cycle: the heat addition process, 2-*a*-3, is under constant-pressure and the heat rejection process, 4-*b*-1, includes two heat rejection processes. The heat rejection process, 4-*b*, is under constant-volume and the other heat rejection process, *b*-1, is under constant-pressure.

Brayton cycle: the heat addition process, 2-*a*-3, is under constant-pressure and the heat rejection process, 4-*b*-1, is under constant-pressure, as shown in Fig. 5.

Ideal Otto cycle and ideal Brayton cycle have been analyzed in this work. All the other ideal gas power cycles can be analyzed with the general *T-s* diagram, Fig. 6. If only the irreversibilities in the compression and the expansion processes are considered with the isentropic efficiencies, the actual gas power cycles can be represented with the general *T-s* diagram, as shown in Fig. 7. The actual gas power cycles can then be designed and analyzed according to their purposes.

Fig. 7 General *T-s* diagram for the gas power cycles between $T_1$ and $T_3$

# 5, Conclusion remarks

Gas power cycles in the course of Engineering thermodynamics are the important content of practical application of basic theory and principle of Engineering Thermodynamics. The textbooks from home and abroad introduce mainly the composition, the performance parameters (such as the thermal efficiency) and the performance analysis method for the ideal Otto, the Diesel, the Dual, as well as the Brayton cycles. The variation of thermal efficiency with relevant parameters is discussed. However, there is no clear theoretical suggestion on the economic operation parameters of gas power cycles, especially for piston engines, in the textbooks, so the engineering applications can not be well directed. After learning this part of course contents, students are difficult to abandon the inveteracy idea that the most efficient device is the best gas power engine.

Taking engines of ideal Otto cycle and ideal Brayton cycle as examples, this paper analyzes the optimum (economical) compression ratio or pressure ratio at which the specific cycle net work is maximum for an engine. Following the same idea, the optimal compression/pressure ratio of other gas power cycles can also be derived, including ideal and actual cycles (only the irreversibilities of compression and expansion processes are considered, as shown in Fig. 7, and all other processes are reversible processes. There exist the operating conditions of maximum specific cycle net work and maximum efficiency for an actual cycle.), and the operating conditions of maximum specific cycle net work are obtained. According to the purpose of use of the device, the design and the transformation of the device should be based on the maximum cycle specific work (such as movable equipment) or the maximum cycle thermal efficiency (such as fixed equipment), rather than taking the thermal efficiency as the only performance parameter. In engineering applications, this should be the theoretical guidelines.

Through years of university class teaching and engineering practices, the authors realize that the gas power cycle system can not be designed and evaluated only with the cycle thermal efficiency, instead, the largest specific cycle net work should be introduced. The gas power system should operate at the working condition of either the highest thermal efficiency, or the maximum specific cycle net work depending upon its purpose. Designing or modifying or retrofitting gas power devices like so, can really and truly reflect the theoretical guidance of Engineering Thermodynamics to engineering applications. Therefore, it is suggested to add the concepts of maximum cycle specific work and optimal compression ratio (or optimal thermal efficiency) to the teaching process of Engineering Thermodynamics. It is finally suggested to add this part to the existing textbooks when they are revised and republished in the future.

This paper puts forward that in the analyses of gas power cycles, the textbooks of Engineering Thermodynamics should pay more attentions to the two important concepts of the optimal compression ratio, or the pressure ratio, and the maximum specific cycle work, in addition to only its thermal efficiency, so as to realize the clear and correct guidance of the theory and principles of Engineering Thermodynamics to engineering applications and better services to engineering practices.

General *T-s* diagram, as shown in Fig. 6, is proposed for ideal gas power cycles. All the ideal gas power cycles can be analyzed easily with the general *T-s* diagram. If only the irreversibilities in the compression and the expansion processes are considered with the isentropic efficiencies, the actual gas power cycles can be analyzed with the revised general *T-s* diagram, as shown in Fig. 7, according to their purposes.

# 6, References